\RequirePackage{fix-cm}
\documentclass[smallextended]{svjour3}       
\smartqed  
\usepackage{url}
\usepackage{appendix}
\usepackage{amsmath}
\usepackage{graphicx}
\usepackage{lineno}
\usepackage{array}
\usepackage{longtable}
\usepackage{natbib}
\newcommand*\patchAmsMathEnvironmentForLineno[1]{%
\expandafter\let\csname old#1\expandafter\endcsname\csname #1\endcsname
\expandafter\let\csname oldend#1\expandafter\endcsname\csname end#1\endcsname
\renewenvironment{#1}%
{\linenomath\csname old#1\endcsname}%
{\csname oldend#1\endcsname\endlinenomath}}%
\newcommand*\patchBothAmsMathEnvironmentsForLineno[1]{%
\patchAmsMathEnvironmentForLineno{#1}%
\patchAmsMathEnvironmentForLineno{#1*}}%
\AtBeginDocument{%
\patchBothAmsMathEnvironmentsForLineno{equation}%
\patchBothAmsMathEnvironmentsForLineno{align}%
\patchBothAmsMathEnvironmentsForLineno{flalign}%
\patchBothAmsMathEnvironmentsForLineno{alignat}%
\patchBothAmsMathEnvironmentsForLineno{gather}%
\patchBothAmsMathEnvironmentsForLineno{multline}%
}

\usepackage{multirow}
\usepackage{graphicx} 
\usepackage[table,xcdraw]{xcolor}
\usepackage{epsfig} 
\usepackage{subfigure}

\usepackage[mathscr]{euscript}

\usepackage{amsthm} 
\usepackage{amsmath} 
\usepackage{amssymb}  

\begin{document}

\title{
Estimating the State of Epidemics Spreading with Graph Neural Networks
\thanks{$^*$These authors contributed equally to this work.}
}


\author{Abhishek Tomy$^*$ \and
        Matteo Razzanelli$^*$ \and
        Francesco Di Lauro \and
        Daniela Rus \and
        Cosimo Della Santina
}


\institute{Abhishek Tomy \at
              Cognitive Robotics Department, 3ME, TU Delft, Delft, Netherlands.
           \and
           Matteo Razzanelli \at
              Proxima Robotics srl, Pisa 56124, Italy. 
           \and
           Francesco Di Lauro \at
              Big Data Institute, University of Oxford, Oxford, UK.
           \and
           Daniela Rus \at
              Computer Science and Artificial Intelligence Laboratory (CSAIL), Massachusetts Institute of Technology (MIT), 32 Vassar St, Cambridge, MA 02139, United States.
        \and
        Cosimo Della Santina \at 
        Cognitive Robotics Department, 3ME, TU Delft, Delft, Netherlands; \\
        Institute of Robotics and Mechatronics, German Aerospace Center (DLR), Oberpfaffenhofen, Germany. \\
        \email{cosimodellasantina@gmail.com}
}


\maketitle

\begin{abstract}
When an epidemic spreads into a population, it is often unpractical or impossible to have a continuous monitoring of all subjects involved.
As an alternative, algorithmic solutions can be used to infer the state of the whole population from a limited amount of measures.
We analyze the capability of deep neutral networks to solve this challenging task. 
Our proposed architecture is based on Graph Convolutional Neural Networks. As such it can reason on the effect of the underlying social network structure, which is recognized as the main component in the spreading of an epidemic.
We test the proposed architecture with two scenarios modeled on the CoVid-19 pandemic: a generic homogeneous population, and a toy model of Boston metropolitan area.
\end{abstract}

\section{Introduction}
\label{intro}
Many natural and artificial systems can be described with models whose state assumes value on a graph rather than on a standard Euclidean space.
Within this class of systems, the problem of estimating the full state from partial measurements is a very relevant one.
If the network follows linear and continuous dynamics, standard techniques can be used.
Yet, things get substantially more complicated as soon as non\--ideal effects are modeled.
For example, \citep{battistelli2012data} introduces constraints in communications bandwidth.
State estimation for networks with distributed delays is discussed in \citep{liu2008synchronization}. A similar problem is dealt in \citep{wang2005state} for the state estimation of a delayed neural network with known output, and in \citep{xu2017robust} for parameter uncertainty and randomly occurring distributed delays. The case of switched networks with communication constraints is discussed in \citep{zhang2017asynchronous}. 
In this context, much attention has also been devoted to distributed estimation algorithms \citep{soatti2016consensus,ding2019survey}. For example, \citep{liu2017kalman} proposes a consensus-based Kalman filter for sensor networks subjected to random link failures, \citep{ding2017distributed} introduces a distributed filter robust to malicious attacks, and \citep{battistelli2016stability} proposes a distributed extended kalman filter for sensor networks measuring a single nonlinear dynamics.

\begin{figure}[t]
    \centering
    \includegraphics[width = \columnwidth]{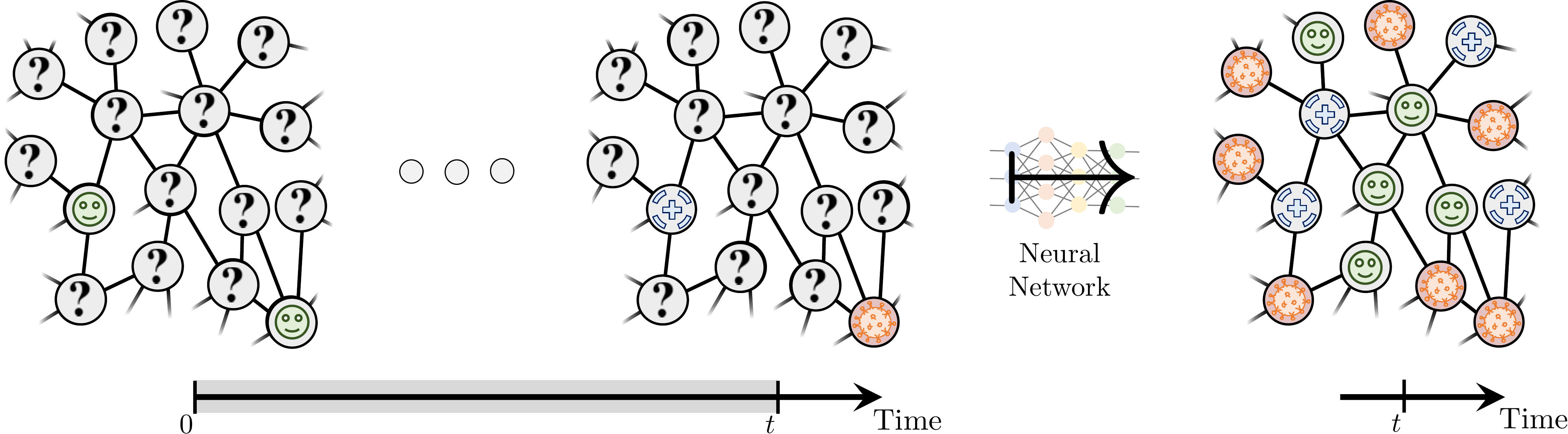}
    \caption{The goal of this work is to test the use a neural architecture to extract the full state of an epidemic spreading on a social network, from the knowledge of the health state evolution of a small set of subjects.}
    \label{fig:task}
\end{figure}

A network dynamics with interesting applications and behavior is the one describing the spreading of an epidemic within a fixed population~\citep{kiss2017mathematics}.  An effective way of modeling this behavior is to describe the social network as a graph. Each node represents either a subject or a group of subjects, and the arcs the contacts. Simple rules are then used to describe the spreading. 
For example, these models have been used to describe the spreading of Covid-19. In \citep{linka2020outbreak} nodes represents European nations. The use of multi\--level networks is discussed in \citep{nande2021dynamics}. 
%
%
A survey on the interplay of diseases, behaviors, and information spreading in epidemics is provided in \citep{wang2019coevolution}.
Network models have been later extended to simplicial complexes in \citep{iacopini2019simplicial}.

Estimating the state of a epidemics from a reduced number of measurements has clear practical implications. For example, being able to estimate not only the number of infected subjects, but also who those infected subjects are, can allow to implement precise isolation policies \citep{bahr2009exploiting,block2020social}, feedback strategies \citep{di2020covid,kompella2020reinforcement}, and possibly prevent the generation of clusters \citep{shim2020transmission}. 
Nonetheless, we are not aware about previous works in epidemiology dealing with this challenge on the subject (i.e. node) level. 
Several works deal instead with the much more common problem of extracting robust statistics on the total amount of subjects being infected, recovered, hospitalized etc \citep{peni2020nonlinear,britton2019stochastic}. This estimation can be used to forecast the evolution of the epidemics \citep{valle2020predicting,tizzoni2012real}. Despite requiring to reason on the network dynamics, the task is still such that it can be attacked with model based techniques, since it is essentially a forward integration. 

Instead, estimating the full state of the epidemics is an essentially more difficult problem since it requires reasoning backward on the effects that the nodes of which we know the state could have had on the unknown states.
This task is made even harder by the highly nonlinear, state\--discrete, and stochastic dynamics which characterizes these systems (see Sec. \ref{sec:dynamic_model}).
%
%
This makes very hard to make inference on the level of the subjects directly using model based techniques.
%
%
In this work we investigate the use of deep learning for creating a nonlinear inference system which can solve the discussed problem \citep{brunton2019data}. 
Recently, many works have dealt with the generalization of deep learning to non Euclidean domains \citep{bronstein2017geometric}. Particular interest have been given to deep learning on graphs \citep{scarselli2008graph,zhou2018graph,bacciu2020gentle}, i.e. to the learning from data of the graph type. Many of these techniques have been chategorized under the umbrella term Graph Neural Networks (GNNs).
%
%
We are interested here in the use of GNNs as classifiers of nodes. The goal is to determine the labeling of nodes by integrating available information on them and on their neighborhood \citep{kipf2016semi}. This is for example used as a recommendation engine - see Pinterest \citep{ying2018graph}, and Uber Eats \citep{ubereats}.
%
%
This task naturally generalizes to the case of state reconstruction, by considering as desired output the full state of the system. 
We apply this strategy to epidemics, by combining multiple GNN layers with a mechanism for codifying temporal information. 
The goal of this work is summarized in Fig. \ref{fig:task}.
We test the results by using state of the art models of epidemics, with particular focus on CoVid-19 spreading in Italy and United States.
%
%
%
Our results show that GNNs can be a viable solution to state reconstruction problem, even when the number of monitored subjects is as low as the $5\%$ of the population.

Note that several works already applied GNNs to epidemics, specifically in the CoVid-19 context. Yet the focus has been different w.r.t. the present work. In \citep{kapoor2020examining,gao2020stan} graph neural networks are used to forecast the pandemic evolution. An inverse problem is instead tackled in \citep{cutura2020deep}, where authors deal with the temporal reconstruction of the epidemics spreading. Similarly, in \citep{shah2020finding} these techniques are used to identify the patient zero.

\section{Epidemics on Networks}
\label{sec:dynamic_model}
\begin{figure}[!t]
    \centering
    \begin{minipage}{.65\textwidth}
        \centering
        \includegraphics[width=0.45\linewidth,trim = 5cm 11cm 5.5cm 10cm, clip]{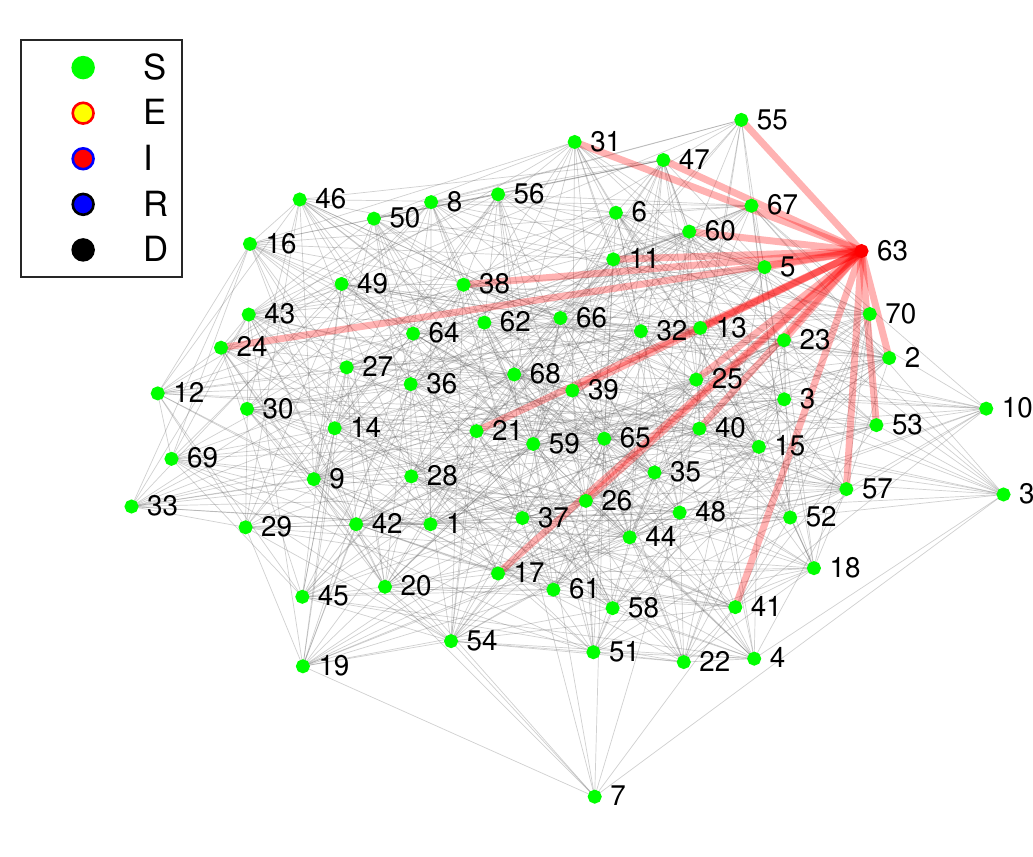}
        \includegraphics[width=0.45\linewidth,trim = 5cm 11cm 5.5cm 10cm, clip]{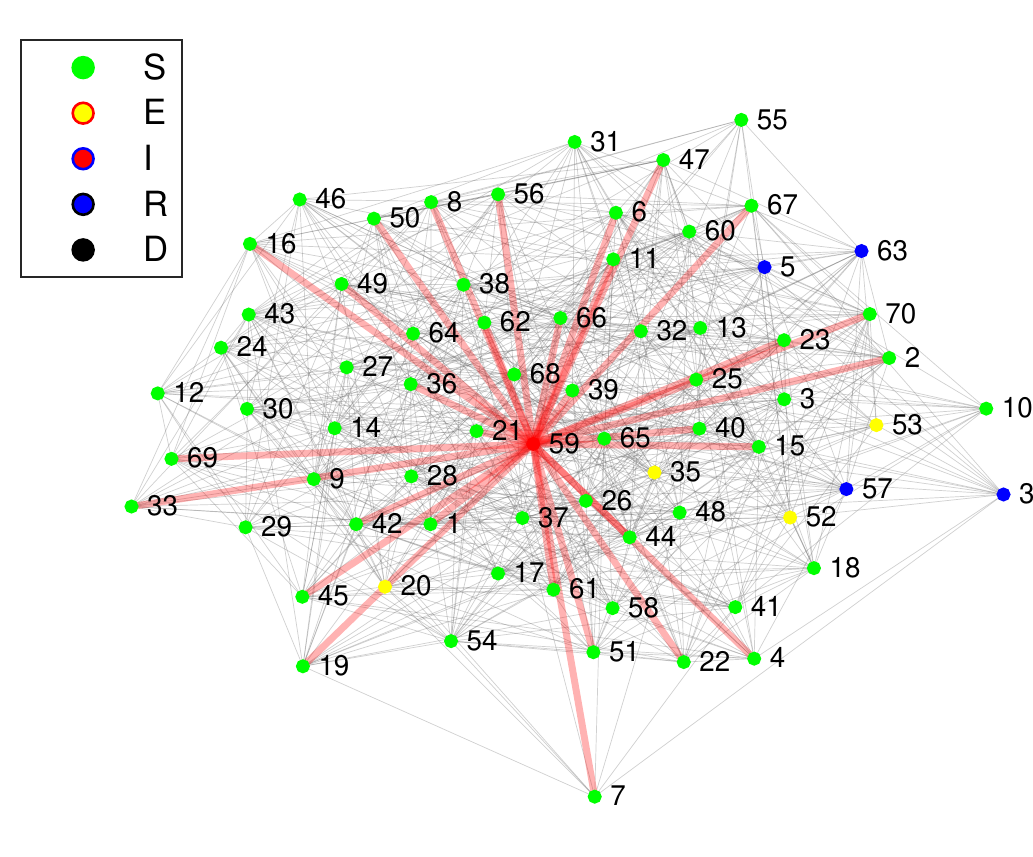}
        \includegraphics[width=0.45\linewidth,trim = 5cm 11cm 5.5cm 10cm, clip]{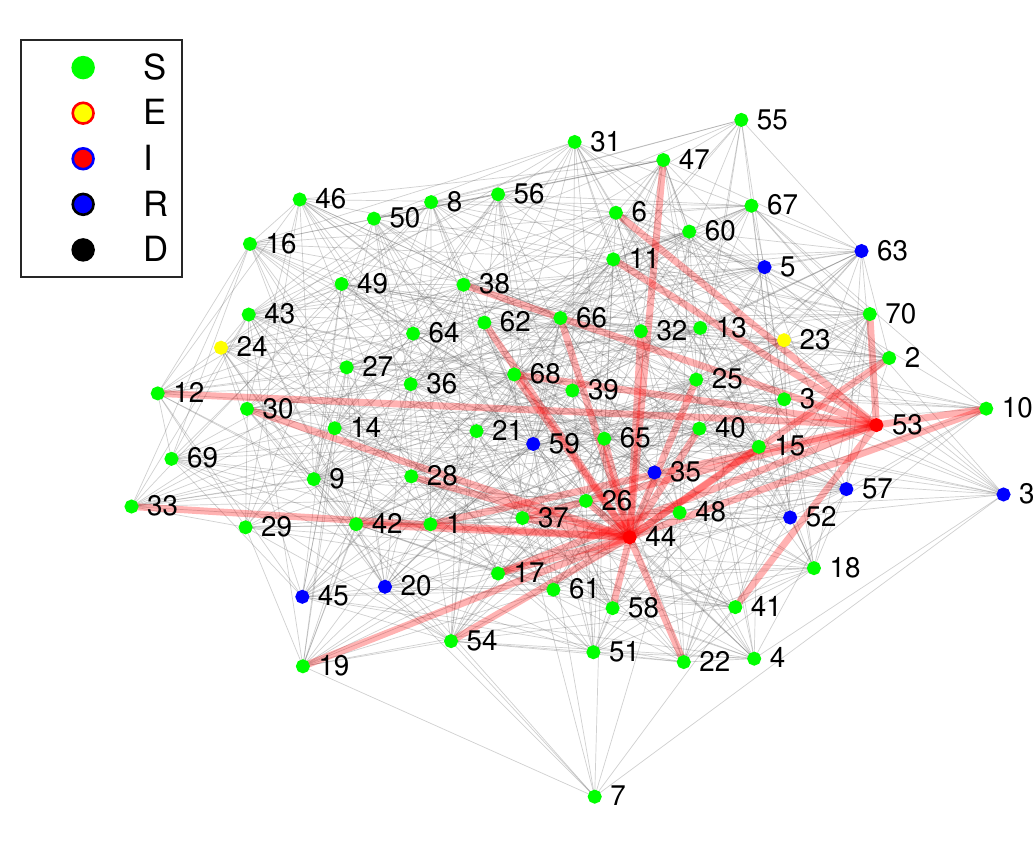}
        \includegraphics[width=0.45\linewidth,trim = 5cm 11cm 5.5cm 10cm, clip]{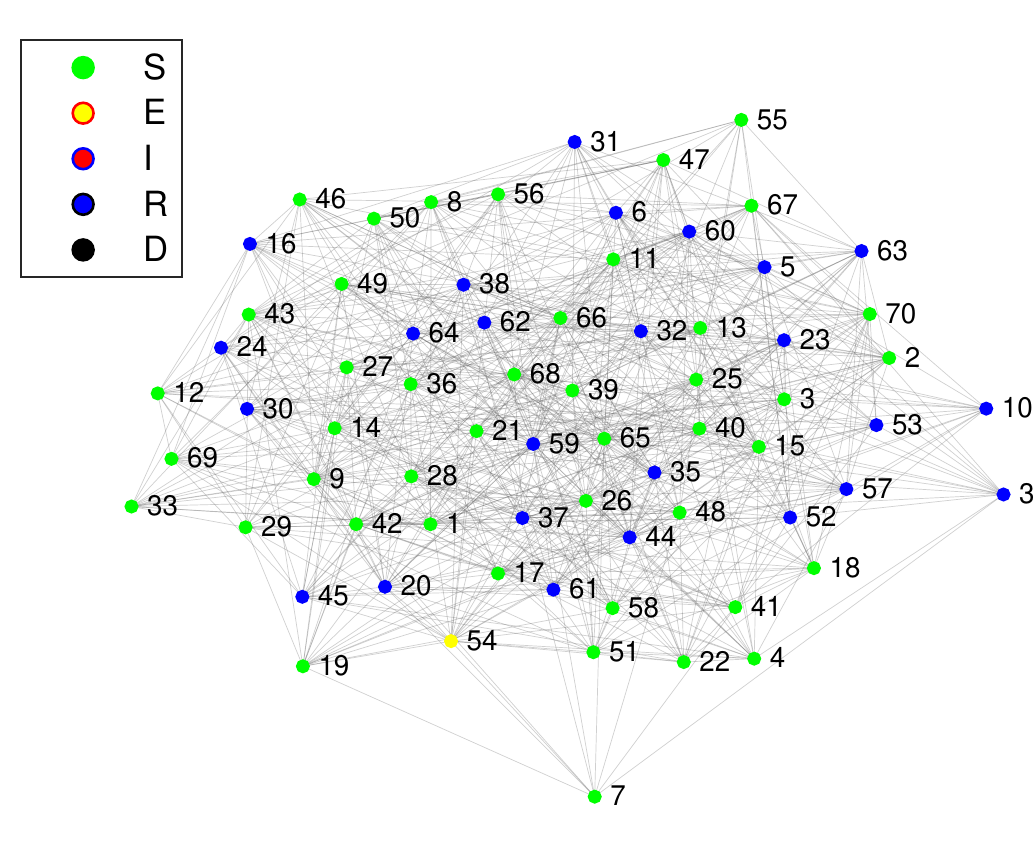} 
    \end{minipage}%
    \begin{minipage}{0.35\textwidth}
        \centering
        \includegraphics[width=0.95\textwidth]{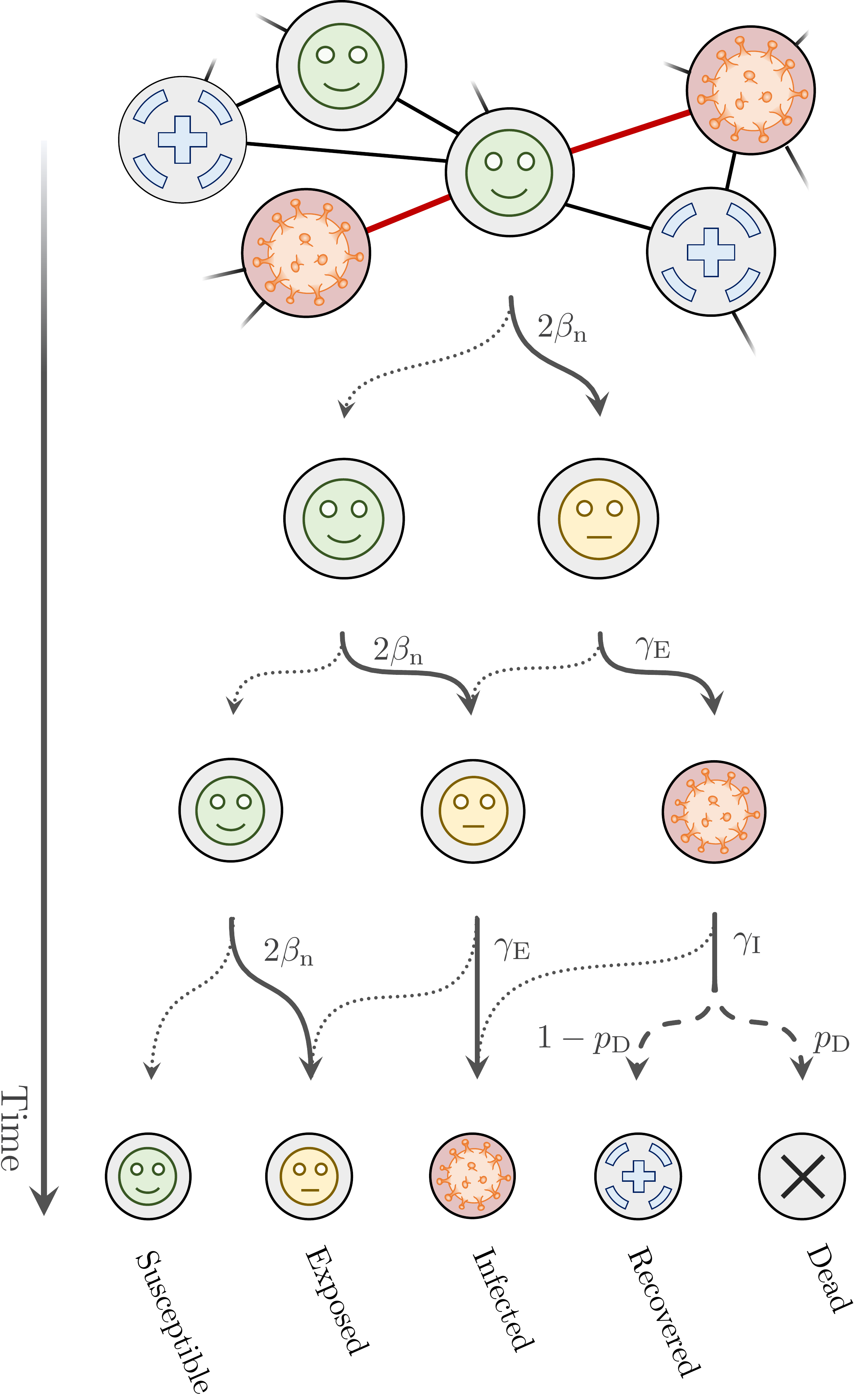}
    \end{minipage}
    \caption{(left) four snapshots of the state of a small Erd\H{o}s-R\'enyi network of size $70$ with average degree $20$.Links in red carry the infection from an infectious node to a susceptible neighbor. (right).  Representation of the transitions a node might undergo with the epidemic model.}
        \label{fig:SEIRD}
\end{figure}
Deterministic models for epidemic spreading on a population are well-known because of their simplicity and their usefulness in terms of giving good prediction in terms of aggregate statistics (such as the total number of infected nodes) of the population.
Yet, these models do not allow to describe the actual spreading of the epidemics on a population, thus preventing the implementation of targeted measures.
For our objective, we need a model that can capture the fact that each individual is part of a social structure, and that the intrinsic hazard of getting infected depends not only on how many people they interact with,  but also on how far they are from clusters of infections.  A natural candidate 
is the framework of Network Epidemiology~\citep{PastorSatorras2015,kiss2017mathematics}. This framework allows to separate the topological properties of a contact network from the biological dynamics of the disease progression.

\subsection{Network Model}
A network is described as a set $(V,\mathcal{E})$, where $V$ is a set of $N$ nodes (or vertices), and $\mathcal{E}$ is a set of edges (or links) connecting nodes, i.e. tuples $\{u,v\}$, where $u,v \in V$. In terms of modeling, individuals are associated with nodes, and contacts that are at risk of carrying the disease as links between nodes. For simplicity, we consider undirected networks, such that $\{u,v\} \in \mathcal{E} \iff \{v,u\} \in \mathcal{E}$. Figs. \ref{fig:SEIRD} shows a pictorial representation of network.
%

\subsection{Epidemic model on Network}
We consider a model for disease transmission inspired by recent modeling of Covid-19~\cite{yang2020modified, he2020seir}. 
Each individual is in one of the following states: $S$ (susceptible), $E$ (exposed), $I$ (infected/infectious), $R$ (recovered), or $D$ (deceased). For this reason, this model is known as SEIRD.
Fig.~\ref{fig:SEIRD} illustrates  the possible transitions of a susceptible node that is in contact with two infectious neighbors.
Outbreaks are modeled as Markovian processes on the generated network, in which an infected node spreads the disease, via links, to its susceptible  neighbors  at a constant rate $\beta$,  turning them into exposed. Exposed nodes represent people who are undergoing their latent people, and are about to become infectious. The next transitions that exposed nodes undergo are network-independent. An $E$ node becomes $I$ after a time exponentially distributed with rate $\gamma_E$. Once a node is infectious, he transmits the disease to its neighborhs at a constant rate $\beta$. The node eventually stops being infectious after an exponentially distributed random time with rate $\gamma_I$.  When this happens, with probability $p_{\mathrm{D}}$ the node becomes $D$ - representing individuals that do not survive to the disease. The remaining nodes are instead recovered and play no further role in the epidemic.  At time $t=0$, $I(0) = N \imath(0) \ll N$ randomly chosen nodes are infected. The remaining ones are initialized as susceptible. 
We use a Gillespie algorithm~\citep{Gillespie1977} adapted to networks~\citep{kiss2017mathematics} to simulate this process. In Fig~\ref{fig:SEIRD} we show a realization of an outbreak on a network of modest size, to highlight how the topology impacts the dynamics.


We describe the the evolution of the state of the pandemic on the network as
\begin{equation}
    x:\mathbb{R}^+ \rightarrow \{S,E,I,R,D\}^N.
\end{equation} 
Therefore at each time $t > 0$ the variable $x(t)$ provides a full picture of thee spreading of the disease. Without loss of generality, we consider $t$ to be expressed in days. 
We refer to the state of the node $i \in V$ as $x_i \in \{S,E,I,R,D\}$. 

\section{State inference from incomplete data}
\label{sec:prob_statement}


\subsection{Goal}\label{sec:goal}

Consider the graph $(V,\mathcal{E})$ describing the social network. We hypothesize to have full knowledge of the state of a subset of nodes $\mathcal{M} \subset V$ at the end of each day.
We will populate $\mathcal{M}$ by selecting nodes from $V$ according to an uniform random distribution.
We therefore define the set of measurements as $y \in \{S,E,I,R,D\}^{\#\mathcal{M}}$.
Finally, for the prediction purposes, classes are combined based on their usefulness in intervention into 3 classes.
Our goal is thus to find an algorithm which implements the following mapping
\begin{equation}\label{eq:general_nn}
    \{V,\mathcal{M},\mathcal{E},y([0,t))\} \mapsto \Tilde{x}(t)
\end{equation}
where $\Tilde{x} \in \{S, E+I, R+D\}^N$ is our reconstructed state, with $E+I$ representing nodes that either exposed or infected, and with $R+D$
We want \eqref{eq:general_nn} to be such that $\Tilde{x}$ is as coherent as possible with the full state $x$.
Indeed, in the practice we are only interested in knowing if the subject is healthy ($S$), has contracted the virus ($E$ and $I$), or is no more infected ($R$ and $D$).
This challenge is summarize in Fig. \ref{fig:task}.

Note that in this work we assume full knowledge of the social structure  $(V,\mathcal{E})$ as input for the network. This is a strong assumption that we will relax in future work.
Also, we will discuss the robustness of the algorithm to changes of topology.



\subsection{First stages}
\begin{figure}[t]
    \centering
    \includegraphics[width = \columnwidth]{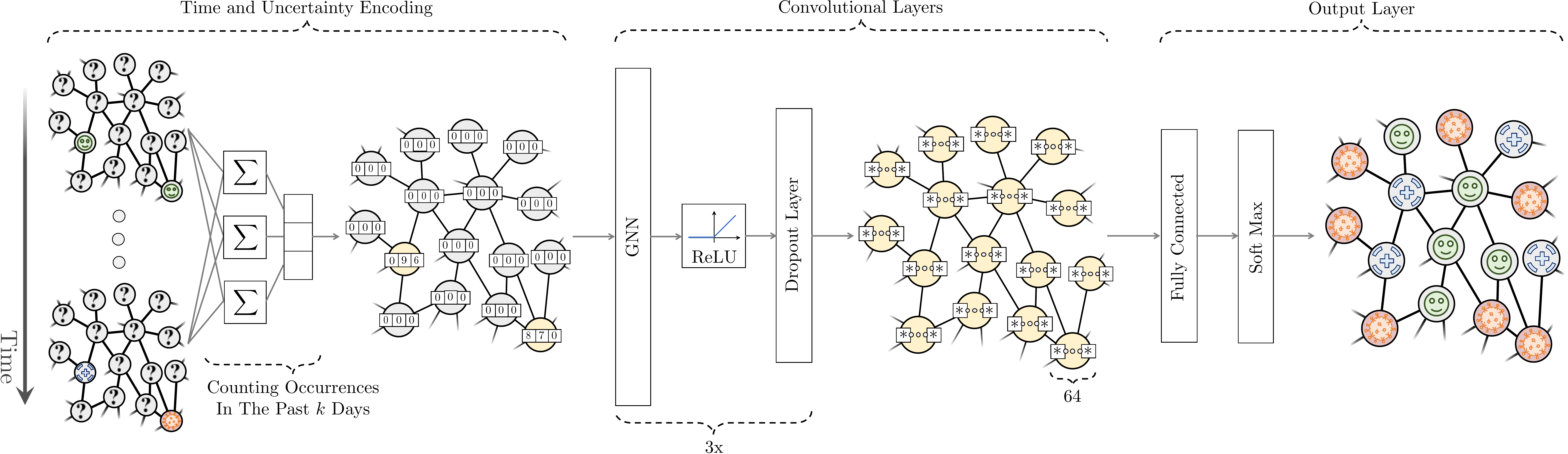}
    \caption{The proposed architecture is made up of three stages. The first one samples data from the evolution of the known nodes $y$ in the past $k$ days, and counts the occurrences of the three classes. In this way it creates labels $l$ encoding the temporal information. The second stage performs most of the computation, and it is made of three graph convolutional layers. Finally, the high dimensional internal information is compressed again by the output layer, i.e. a fully connected network and a soft max. The output is an estimation of the current full state of the epidemics spreading.}
    \label{fig:GNN}
\end{figure}

We start by transforming $y([0,t))$ in a data structure that can be effectively put as input of our neural network.
More specifically, we introduce the nodes label $l \in \mathbb{N}^{N \times 3}$. 
For all $i \in \mathcal{M}$, the vector $l_i$ codifies the state of the nodes in the past $k$ days. We do that in a bag\--of\--words fashion \citep{weinberger2009feature}. 
We sample $y$ on a daily basis $y_i(\lfloor t\rfloor), y_i(\lfloor t\rfloor - 1) \dots, y_i(\lfloor t\rfloor - k + 1)$.
The value $k \in \mathbb{N}$ is an hyperparameter which will be later optimized.
We then take $l_{i,1}$ equal to the number of times the state $S$ appears in the sampling. Similarly, $l_{i,2}$ counts the occurrences of $E$ and $I$, and $l_{i,3}$ of $R$ and $D$. Therefore the sum of elements in $l_i$ is always equal to $k$ for $i \in \mathcal{M}$.
The remaining nodes are labeled as \textit{unknown} by taking $l_i = 0$ for all $i \notin \mathcal{M}$.
These operations are graphically summarized in the left part of Fig. \ref{fig:GNN}.

\subsection{Neural architecture}
%
Graph Neural Networks 
operate in the domain of the graph. In the graph, each node comes with its label. A common framework in the GNN is the classification problem setup where the goal is to predict the label of the unlabeled nodes given the labeled ones. As mentioned before we want to predict the full state of the pandemics spreading $x$, see Fig. \ref{fig:task}.

The central part of Fig. \ref{fig:GNN} shows the core GNN layers in our architecture. The target of our GNN is to learn the state embedding $l_i\in\mathbb{N}^3$ for $i=1,2,3$, which contains the information of neighborhood for each node. The initial node feature corresponds to the node state itself, encoded in binary vector $\in \mathbb{N}^3$ that contains only one element equal to 1. We preprocess this information by integrating $l_i$ along the time horizon of $k$. We cannot use $k$ too big  to avoid that the neural network leverages on this pattern to recognize that the state coincide with the node label.
Due to the fact that it is a classification problem setup, we then mask a certain percentage of node ($95-90-80\%$) depending on the scenario we are considering.
We finalize the preprocess by loading data by batch by using the Dataloader class defined inside the Pytorch library \citep{paszke2019pytorch}. Thanks to a specific variable, named 'batch', the data loader can associates node and edges to a specific graph. Since a DataLoader aggregates nodes, edges and the features from different graphs into batches during the message passing layers, the GNN model needs this information to know which nodes belong to the same graph.
For what concern message passing layer, it describes how $l_i$ is passed through the layers of the network to create the node embedding. As we know, the message passing layer is the result of the generalization of the convolution operator by extending the concept of the neighbourhood from pixels to nodes \citep{kipf2016semi}.
Given the state of the node $i$ at the layer $h$, $l_i^{h}$ we find the $l_i^{h+1}$ by applying the activation function of the message passing layer to $l_i^{h}$ and the aggregation of $l_j^{h}$ where $j\in\mathcal{N}_i$ is a neighbour of node i (and $\mathcal{N}_i$ is the neighbourhood of node $i$). As the node embedding evolve through the message passing layers, as the knowledge of the neighborhood of each single node increases.
Thus, the message passing layers enlarge, in general, the size of the node feature. The number of the message passing layers could be considered again as a hyperparameter.
Without loss of generality, in our case, three message passing layers with a rectifier as the activation function (ReLu) are considered. The first message passing layer has an input size of 3 (i.e. the number of features), and output size of 64. The second and the third message passing layers have an input and output sizes of dimension 64. Between layers there the dropout regularization method is used during training to avoid over-fitting.

As a result of this processing, each node is equipped with a rich description of its possible state as inferenced by its neighbours own representations. This state need to be converted into one of the three states $\{S,E+I,R+D\}$. This is done through the Output Layer (right part of Fig. \ref{fig:GNN}).
First we have a fully connected layer. Its input size is 64 and output size of 3. It is defined with a linear activation function. Then, a $softmax$ function in introduced as defined in the Pytorch library. It is  applied to the observation $l_i$ so to retrieve the highest probability that the node will be labeled with a certain class.

For what concerns the training, an Adam optimizer with a fixed learning rate is defined and we select loss $L_1(\cdot)$ as the cross entropy. Given the unbalanced classes $c\in\{S,E+I,R+D\}$ we compute weight $w_{c}$ to normalize observation $l_i$.
The weight of each class is determined by the $\frac{N_{\mathrm{max}}}{N_c}$ where $N_{\mathrm{max}}$ is the the number of observations in the class with maximum occurrence and $N_c$ is the number of observations or nodes belonging to class $c$ in the training set.
We use the loss function for measuring the performance of the algorithm as described by the Pytorch library.
%
%
The losses are then averaged across observations:
\begin{equation}
    L_1(\cdot)=\frac{\sum_{c=1}^{3} \mathrm{loss}(x_{c})}{\sum_{c=1}^{3}w_{c}}.
\end{equation}
Given a fixed number of epochs (250 in our case) we train our network and we measure the loss function as previously defined to measure the loss.
Hyperparameter optimization is done using balanced accuracy.
Balanced accuracy is calculated as the average of the proportion corrects of each class individually.
Balanced accuracy is suitable for datasets with class imbalance unlike other metrics which may favour results from the majority class.

%

\section{Simulations}
%

We test the proposed architecture in two scenarios, with different topological characteristics. The first one is an homogeneous network, in which any node has the same probability of being connected with all the others. We use this scenario to test extensively the effectiveness and scalability of the method. The second scenario is instrumental to test the neural architecture in a more challenging setting, closer to a real world scenario.

\subsection{Scenario 1: random network}

We consider Erd\H{o}s-R\'enyi networks, which are a class of well-known network models. Such random networks are relatively simple to describe, and at the same time offer some heterogeneity in terms of the degree distribution. The generative algorithm can be described as follows: we start with $N$ isolated nodes, then we place a link between any two nodes with probability $0<p<1$. The degree distribution of the network is therefore binomial $\mathbb{B}(N,p)$. We showcase results for networks with average degree $\langle k \rangle = 30$. This value is comparable with the number of daily contacts at risk as measured in a recent survey~\citep{Melegaro2011}. 
%

We generate training set from 80 realizations, each one happening on a different and randomly generate social network with a population of 500 nodes. 
The epidemic spreads between 0 and 120 days. Yet in the initial month, the behavior is quite stationary due to the well\--know slow increase of the total number of infected subjects. Therefore, a few samples from the initial days is enough to learn the pattern during that period. Only 3 random days are selected from the first month of each realization. All the remaining days from 30 to 120 are used for training.
%
%
The hyperparameters are $0.3$ for dropout, $64$ hidden units, $3$ layers. We use a learning rate of $0.0002$, we train for $250$ epochs, with a batch size of $256$.

\begin{table}[t]\label{tab:scenario_1_500}
\caption{Accuracy (Ac.) and precision (Pr.) of the classification for Scenario 1, evaluated only on the nodes which are not in $\mathcal{M}$. The testing set is generated with networks of $500$ nodes. Three levels of supervision are considered - i.e., number of nodes of $V$ which are in $\mathcal{M}$ as well.}
\begin{center}
\begin{tabular}{ |c|c|c|c|c|c|c|c| } 
\hline
 & Ac., $5\%$ & Pr., $5\%$ & Ac., $10\%$ & Pr., $10\%$ & Ac., $20\%$ & Pr., $20\%$ \\
\hline
$S$        & 0.93 & 0.84 & 0.93 & 0.84 & 0.93 & 0.85 \\ 
$E \!+\! I$& 0.52 & 0.56 & 0.51 & 0.57 & 0.51 & 0.57 \\ 
$R \!+\! D$& 0.74 & 0.78 & 0.75 & 0.77 & 0.76 & 0.77 \\ 
All        & 0.75 & -    & 0.75 & -    & 0.76 & -   \\ 
\hline
\end{tabular}
\end{center}
\end{table}

\begin{figure}[t]
    \centering
    \subfigure[S, 5\%]{\includegraphics[trim = {6cm 16.5cm 6cm 4.5cm}, clip,height = 0.225\columnwidth]{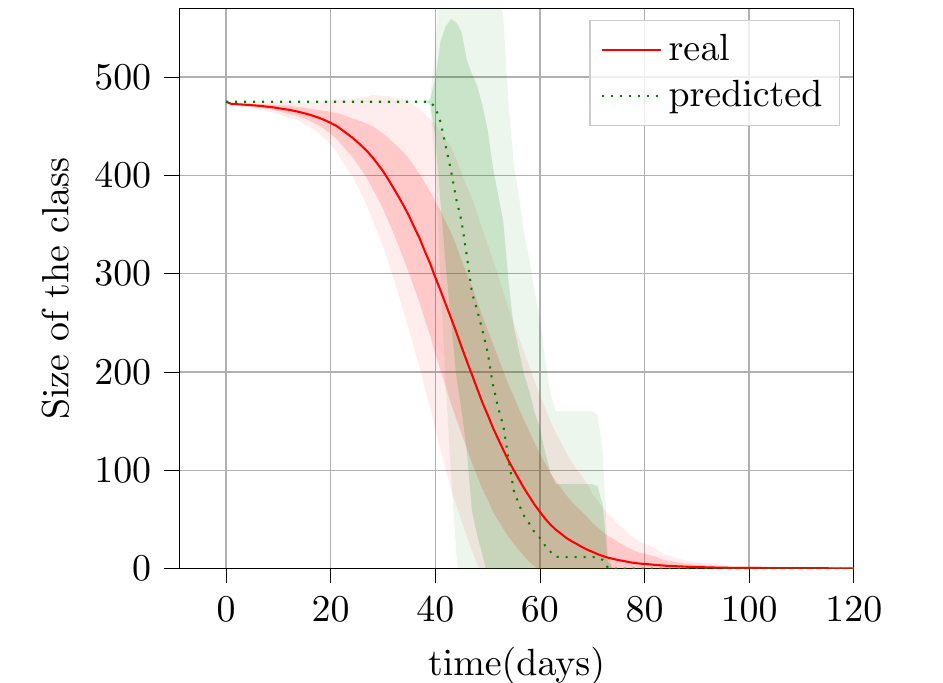}}
    \subfigure[E+I, 5\%]{\includegraphics[trim = {6cm 16.5cm 6cm 4.5cm}, clip,height = 0.225\columnwidth]{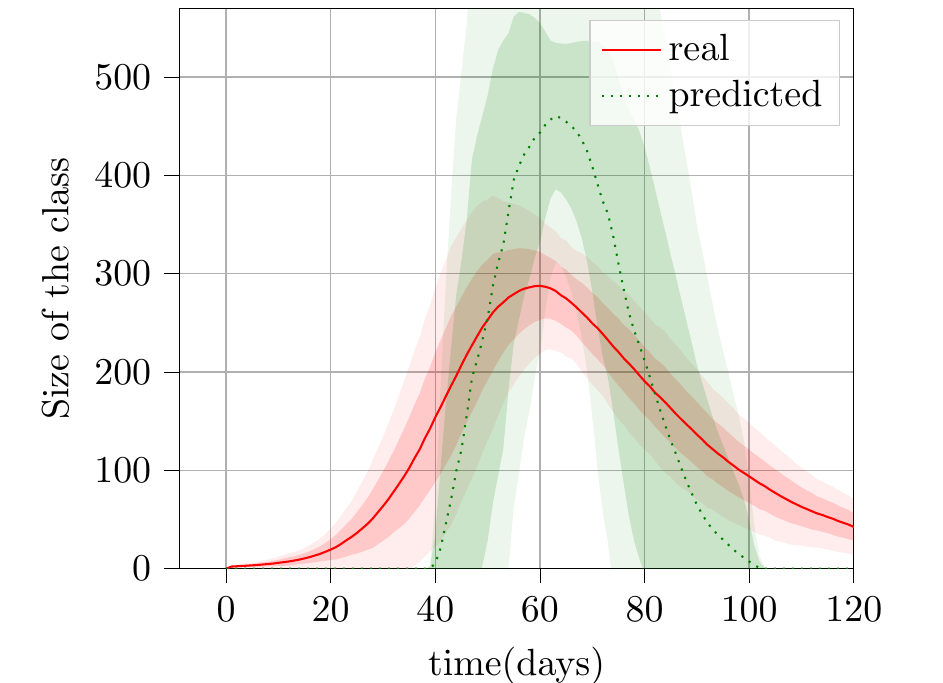}}
    \subfigure[R+D, 5\%]{\includegraphics[trim = {6cm 16.5cm 6cm 4.5cm}, clip,height = 0.225\columnwidth]{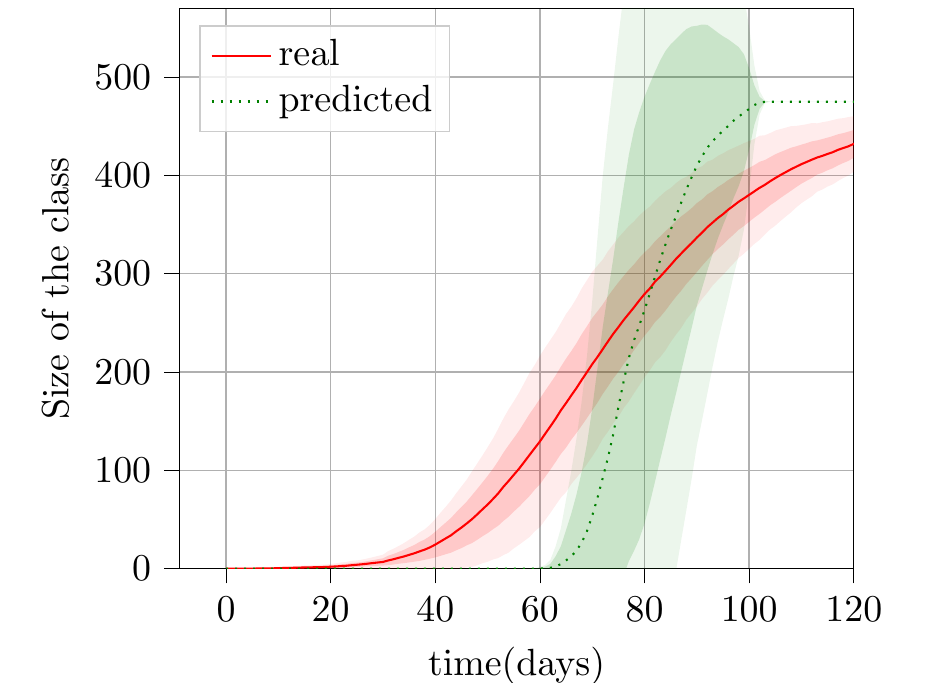}}
    \subfigure[S, 20\%]{\includegraphics[trim = {6cm 16.5cm 6cm 4.5cm}, clip,height = 0.225\columnwidth]{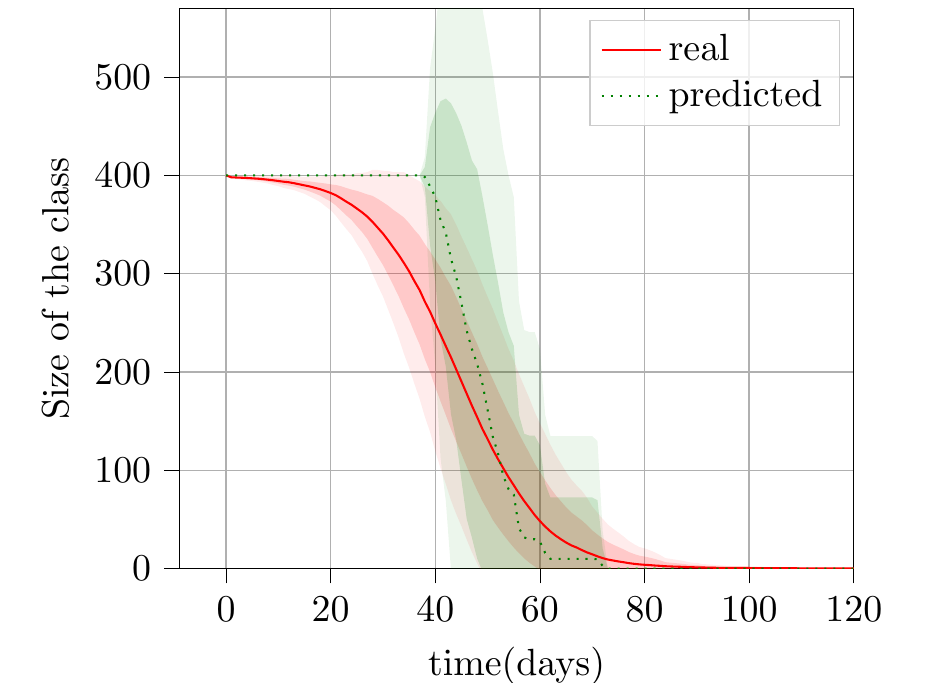}}
    \subfigure[E+I, 20\%]{\includegraphics[trim = {6cm 16.5cm 6cm 4.5cm}, clip,height = 0.225\columnwidth]{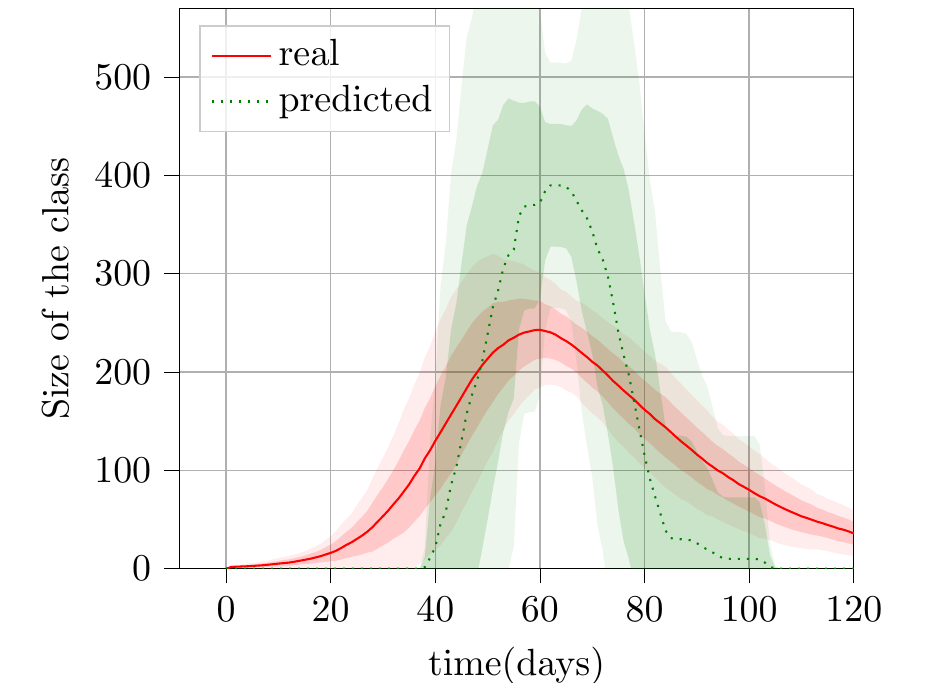}}
    \subfigure[R+D, 20\%]{\includegraphics[trim = {6cm 16.5cm 6cm 4.5cm}, clip,height = 0.225\columnwidth]{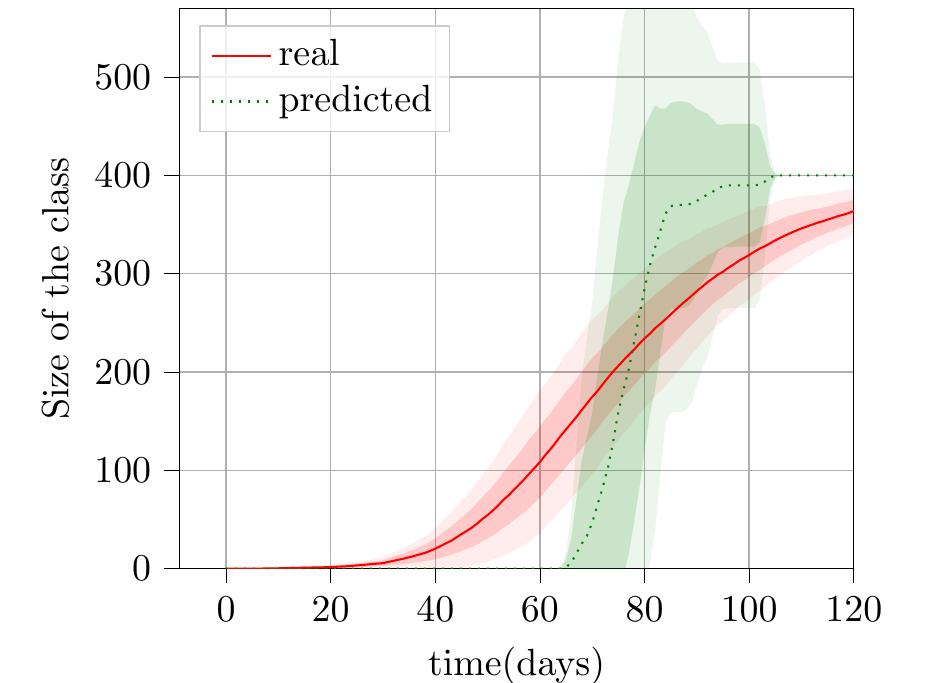}}
    %
    \caption{Evolution of overall statistics associated to the epidemics, when evolving on a medium\--small homogeneous network (scenario 1).
    More specifically, we show the amount of nodes which are susceptible - Panels (a,d) - which got infected by the pathogen - Panels (b,e) - and which either recovered or died - Panels (c,f).
    Actual evolutions are in red, while estimations are in green. The solid lines represent the mean, while the translucent areas the variance.
    Training and testing sets are made of realizations produced by simulating the epidemic spreading on random social networks of 500 subjects.
    In Panels (a-c) only $5\%$ of subjects is tested at any time, while in Panels (d-f) this number reaches $20\%$.
    }
    \label{fig:results_scenario_1_500}
\end{figure}

At first, we test the trained architecture on a set of $40$ realizations, representing evolutions on randomly generated social networks with $500$ nodes (same size of the training set).
We repeat the analysis for the cases in which the size of $\mathcal{M}$ (i.e. monitored subjects) is 5\%, 10\%, 20\% of the size of $V$ (i.e. the total amount of subjects in the considered population).
It is worth to notice that this is a very sparse amount of information. Indeed, $10\%$ of tests with an average connectivity of $10$ means that any node has on average just a single neighbor whose the state is known (see Fig. \ref{fig:task} to get a visual sense of this ratio).
Accuracy and precision of the predictions are provide in Tab. \ref{tab:scenario_1_500}. Note that these values are evaluated only on the nodes which are not part of $\mathcal{M}$ since they are always perfectly known. Thus, we prefer to leave them out to not artificially increase the performance of the neural network. 
Interestingly, the quality of the predictions do not change significantly with the size of $\mathcal{M}$.
In general classes with larger amount of subjects have better performance. This can be due to the higher amount of examples which are available from the training set.
Overall the performance is satisfactory, with a general accuracy always higher than $0.75$.
To get a sense of how these results reflect in the estimation of cumulative statistics of the pandemic evolution, in Fig. \ref{fig:results_scenario_1_500} we plot the total size of each class against the amount of nodes which are classified to be part of that class. The match is good. The network is not sensitive to small deviations of $S$ and $R+D$ from the maximum and the minimum value. This may be due to the fact that so small variations may not be captured by changes in $\mathcal{M}$. Also, the neural network tends to over estimate the presence of subjects which got infected at the pick.
It is very important to stress here that these overall statistics serve here only to get a sense of the overall quality of the network predictions. The goal of the neural architecture is indeed not to estimate these values directly, but the exact way in which each class is spread over the social network. This is an important distinction because the direct estimations of the size of the three classes is a relatively simple task, as discussed in the Introduction.

\begin{table}[t]\label{tab:scenario_1_1e5}
\caption{Accuracy (Ac.) and precision (Pr.) of the classification for Scenario 1, evaluated only on the nodes which are not in $\mathcal{M}$. The testing set is generated with networks of $10^5$ nodes. Three levels of supervision are considered - i.e., number of nodes of $V$ which are in $\mathcal{M}$ as well.}
\begin{center}
\begin{tabular}{ |c|c|c|c|c|c|c|c| } 
\hline
 & Ac., $5\%$ & Pr., $5\%$ & Ac., $10\%$ & Pr., $10\%$ & Ac., $20\%$ & Pr., $20\%$ \\
\hline
$S$        & 0.97 & 0.92 & 0.96 & 0.92 & 0.93 & 0.85 \\ 
$E \!+\! I$& 0.50 & 0.58 & 0.50 & 0.57 & 0.51 & 0.57 \\ 
$R \!+\! D$& 0.79 & 0.81 & 0.80 & 0.81 & 0.76 & 0.77 \\ 
All & 0.83 & - & 0.83 & - & 0.83 & - \\ 
\hline
\end{tabular}
\end{center}
\end{table}

\begin{figure}[t]
    \centering
    \subfigure[S, 5\%]{\includegraphics[trim = {6cm 16.5cm 6cm 4.5cm}, clip,height = 0.225\columnwidth]{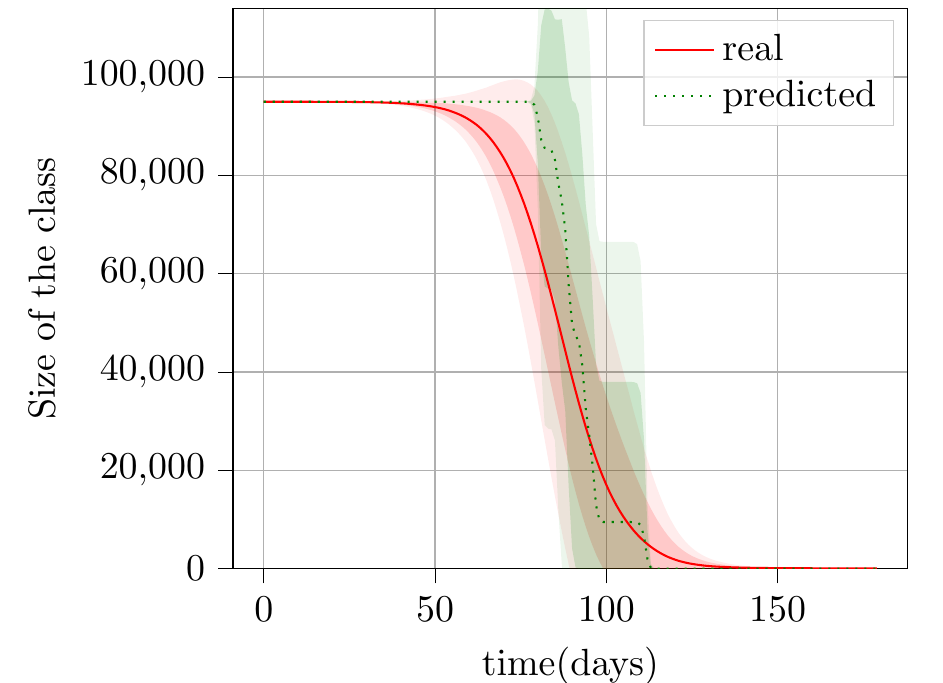}}
    \subfigure[E+I, 5\%]{\includegraphics[trim = {6cm 16.5cm 6cm 4.5cm}, clip,height = 0.225\columnwidth]{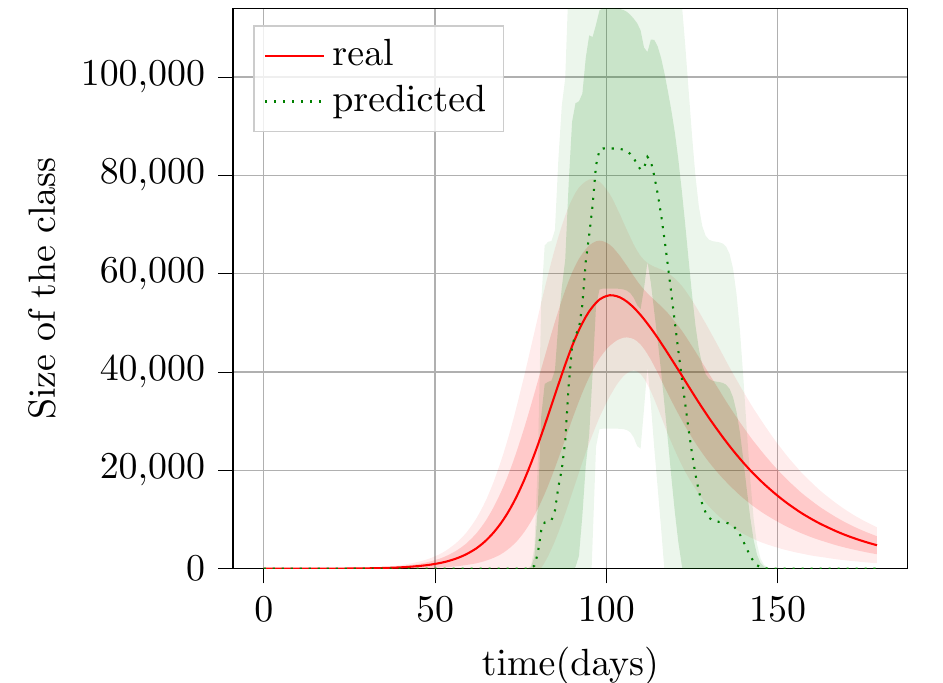}}
    \subfigure[R+D, 5\%]{\includegraphics[trim = {6cm 16.5cm 6cm 4.5cm}, clip,height = 0.225\columnwidth]{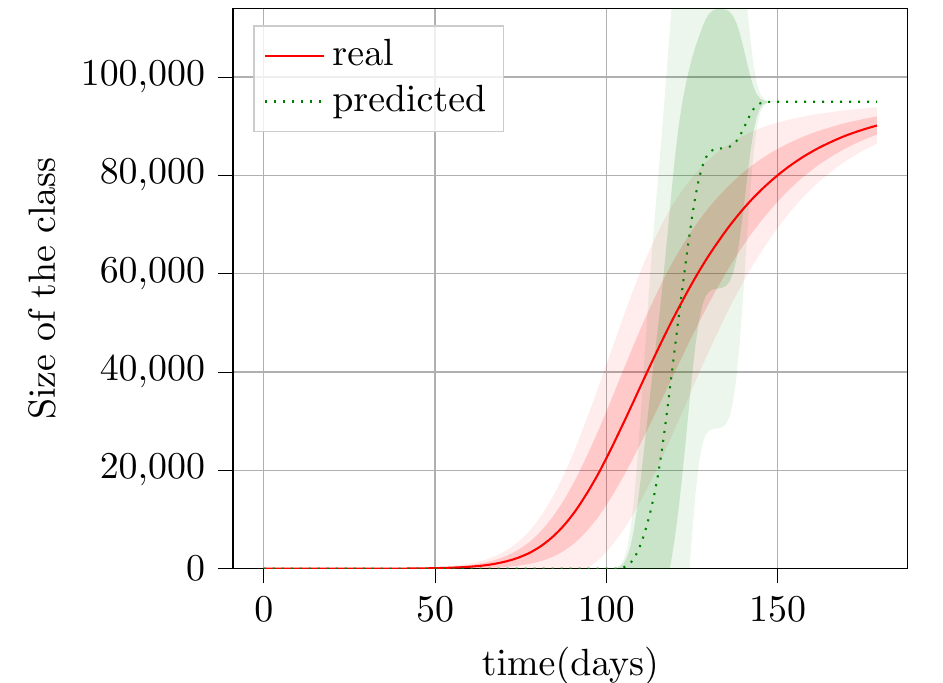}}
    \subfigure[S, 20\%]{\includegraphics[trim = {6cm 16.5cm 6cm 4.5cm}, clip,height = 0.225\columnwidth]{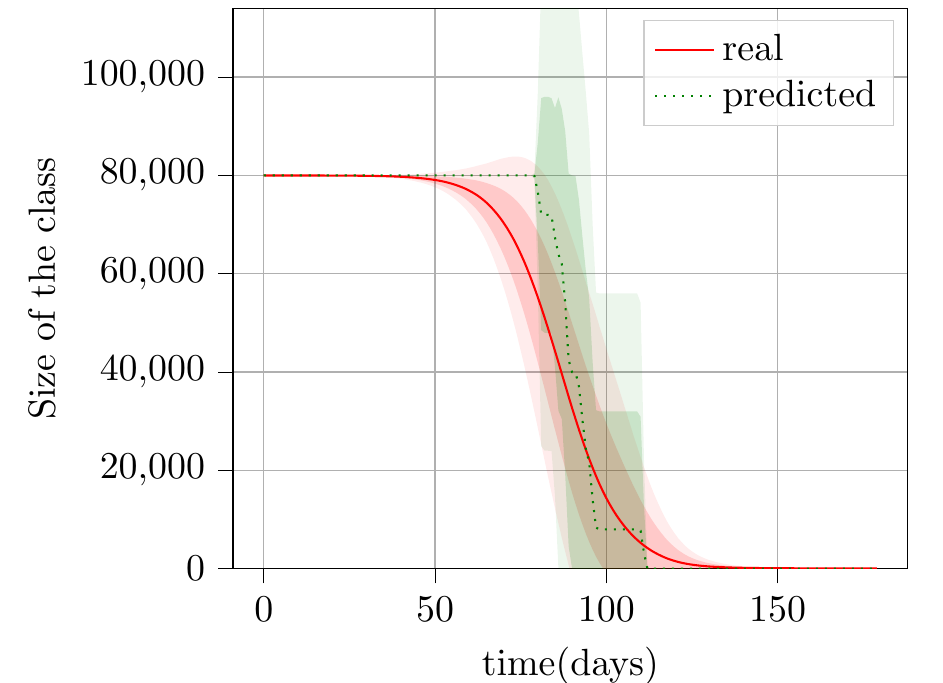}}
    \subfigure[E+I, 20\%]{\includegraphics[trim = {6cm 16.5cm 6cm 4.5cm}, clip,height = 0.225\columnwidth]{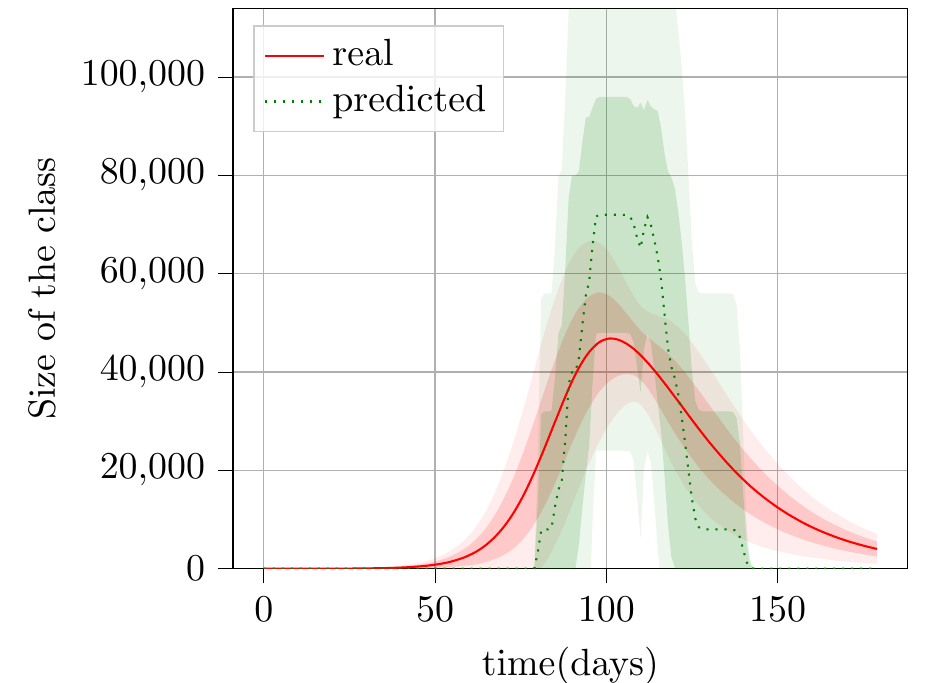}}
    \subfigure[R+D, 20\%]{\includegraphics[trim = {6cm 16.5cm 6cm 4.5cm}, clip,height = 0.225\columnwidth]{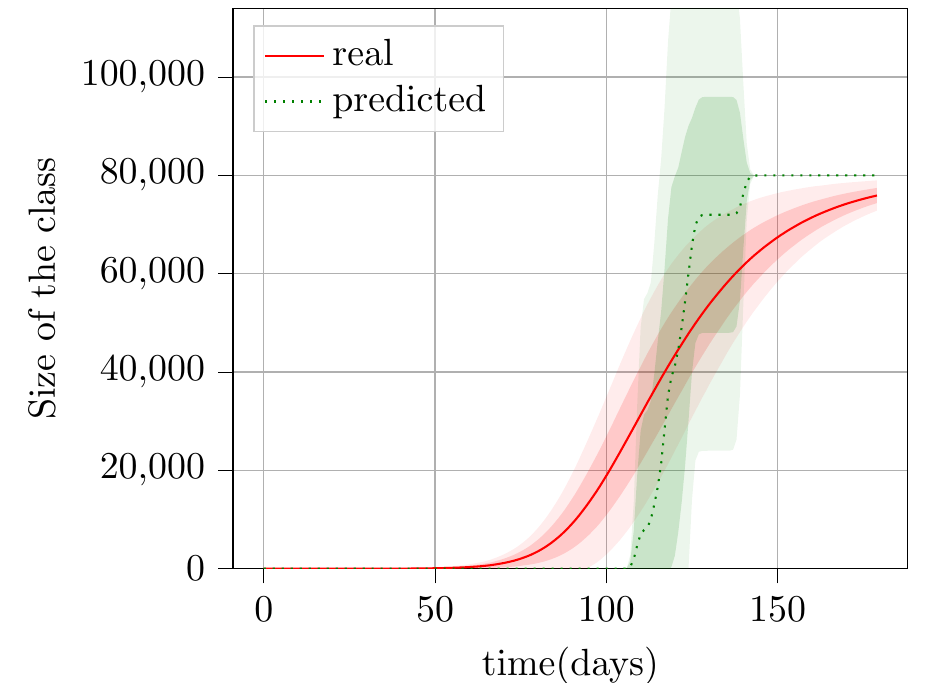}}
    \caption{Evolution of overall statistics associated to the epidemics, when evolving on an homogeneous network (scenario 1).
    Actual evolutions are in red, and estimations in green. The solid lines represent the mean, while the translucent areas the variance.
    The testing set is made of realizations produced by random social networks of $10^5$ subjects. Instead the training set contains only networks which are $500$ nodes big. 
    In Panels (a-c) only $5\%$ of subjects is tested at any time, while in Panels (d-f) this number reaches $20\%$.
    }
    \label{fig:results_scenario_1_1e5}
\end{figure}

A nice property that our architecture inherits from Graph Convolutional Neural network is that once trained it can be applied to graphs of any size.
This is because we directly learn the weights of the convolution operator, which can then be applied to notworks of all sizes. There is however no guarantee that the classification will keep being effective. Indeed, the way in which the pandemic evolves is clearly affected by the size of the social network despite the local rules remaining the same.
We therefore tested the ability of the architecture to generalize to larger populations by building an additional testing set of 10 realizations with a total number of nodes which is several orders of magnitude larger than before: $10^5$ subjects. It is very important to stress that no re\--training is performed. Therefore, we are training the neural architecture with a small\--village community, and testing it with a medium size city.
Tab. \ref{tab:scenario_1_1e5} and Fig. \ref{fig:results_scenario_1_1e5} show the result of this analysis. No essential differences can be observed.
Overall the performance is still satisfactory, with a general accuracy which is even higher than the previous test set and always equal to $0.83$.
This may be due to the fact that larger social networks generate more homogeneous distributions of the illness since the border-effects are less dominant.

\begin{figure}[t]
    \centering
    \includegraphics[width = .8\columnwidth]{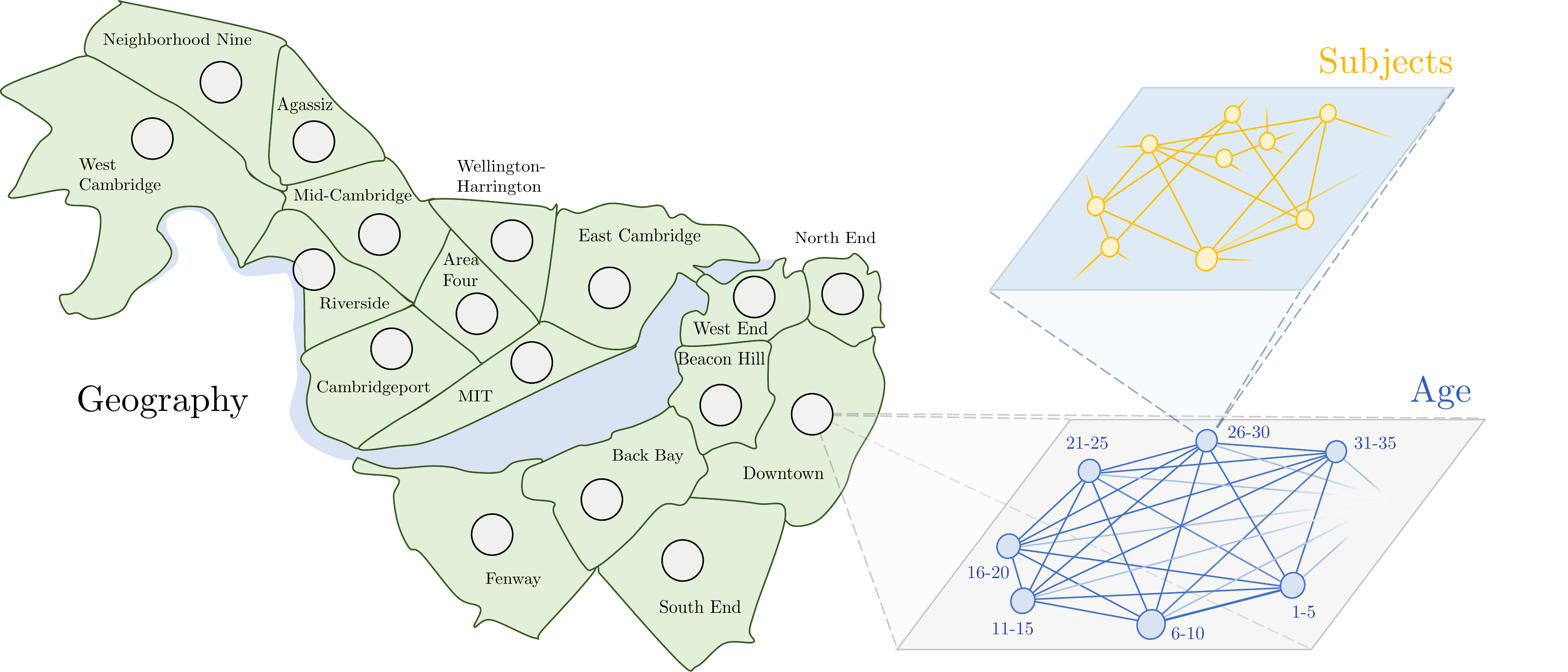}
    \caption{The second scenario is a toy model inspired by the spreading of CoVid in the Boston and Cambridge, Massachusetts. The topology of the graph is built on three layers, which integrate the geographical distribution of the population, the demographics, and the subject level variability.}
    \label{fig:boston_network}
\end{figure}

\subsection{Scenario 2: Boston}

The second scenario we consider aims at modeling a more realistic social structure, such as the one of a relatively big city. We need to consider both a model that takes into account the existence of different neighborhoods and the age distribution of people living in that area. We take as a reference the City of Boston and Cambridge, Massachusetts, USA.

The generative model, which takes inspiration from the work in~\citep{mistry2021inferring}, is divided into three steps,  as in fig.~\ref{fig:boston_network}. Initially, we outline a map of the neighborhoods of the urban area we focus on. At this stage, each neighborhood is a network on its own. The size of each neighborhood is taken from the official website of the city of Boston\footnote{\url{http://www.bostonplans.org/getattachment/7987d9b4-193b-4749-8594-e41f1ae27719}} and Cambridge\footnote{\url{https://www.cambridgema.gov/-/media/Files/CDD/FactsandMaps/profiles/demo_profile_neighborhood_2019.pdf}}. Within each neighborhood, the topology reflects the contact patterns between different age-classes, as described  in the Supplementary material of~\citep{mistry2021inferring}. To do so, we cohort  the population  into age groups of size 5 years, and we model the contact patterns among groups based on their age with a stochastic block  model~\citep{hollandblockmodels}. Stochastic block models are generative models for random graphs that are use to generate topologies that have a community-like structure. Each node is given a unique label (the age cohort). Then, we define a symmetric matrix (known as Affinity matrix) whose elements are $\mathcal{A}_{ij}=p_{ij}$, where $p_{ij}$ is the probability that a node whose label is $i$ is in contact with a node whose label is $j$. The Affinity matrix we use is the Massachusetts age-contact matrix, as described in the supplementary material of~\citep{mistry2021inferring}.  The last step is to connect different neighborhoods by allowing nodes in each neighborhood to have links with nodes from other neighborhoods. To do so, we consider a diffusion-like procedure: for each couple of neighborhoods we place a random number of links between randomly selected nodes from both communities, depending on the length of the shortest path connecting the two on the geographical level: neighborhoods at distance $d$ from each other will share, on average, $1/d$ links with respect to neighborhoods at distance $1$. The number of links shared between any two communities is drawn from a Binomial with probability $p=\frac{1}{50} \frac{1}{d}$. 
%
%

We generated a training composed of 20 realizations, each one happening on a different and randomly generate social network with a population of $10^4$ nodes. This is one order of magnitude less than the actual population of that area. This choice has been imposed by limits on the hardware resources available.
In this scenario, the epidemic spreads over a relatively long period of time, with each day being of importance and different. Hence we select a total of $201$ of days from each realizations, starting $100$ days before the pick of the infection, and ending $100$ days after.
No sample is removed.
The hyperparameters are $0.4$ for dropout, $64$ hidden units, $3$ layers. We use a learning rate of $0.0002$, we train for $250$ epochs, with a batch size of $64$.

\begin{figure}[t]
    \centering
    \subfigure[S, 5\%]{\includegraphics[trim = {6cm 16.5cm 6cm 4.5cm}, clip, height = 0.225\columnwidth]{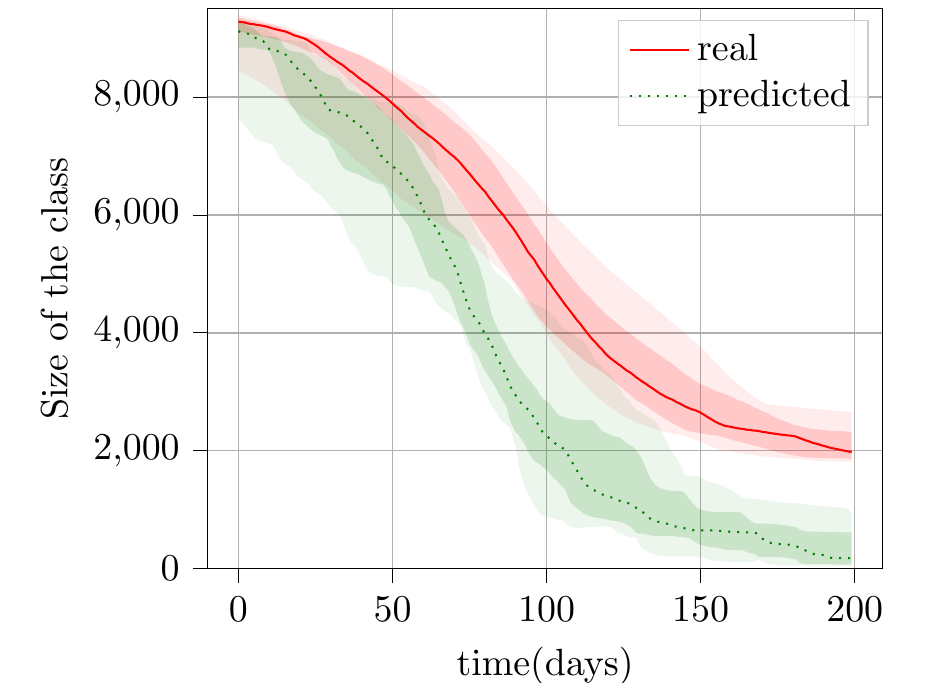}}
    \subfigure[E+I, 5\%]{\includegraphics[trim = {6cm 16.5cm 6cm 4.5cm}, clip, height = 0.225\columnwidth]{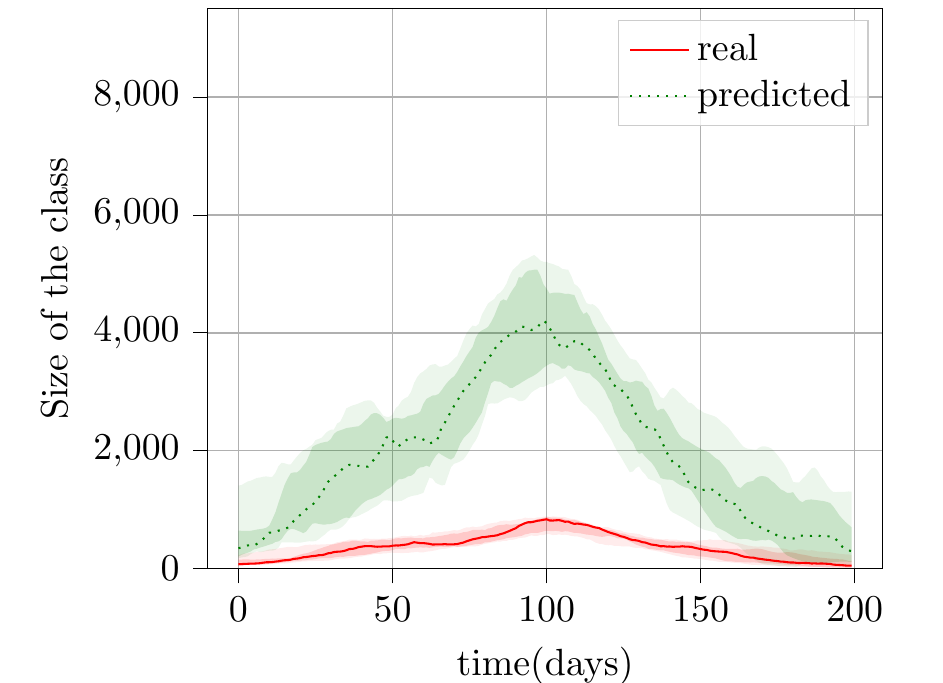}}
    \subfigure[R+D, 5\%]{\includegraphics[trim = {6cm 16.5cm 6cm 4.5cm}, clip, height = 0.225\columnwidth]{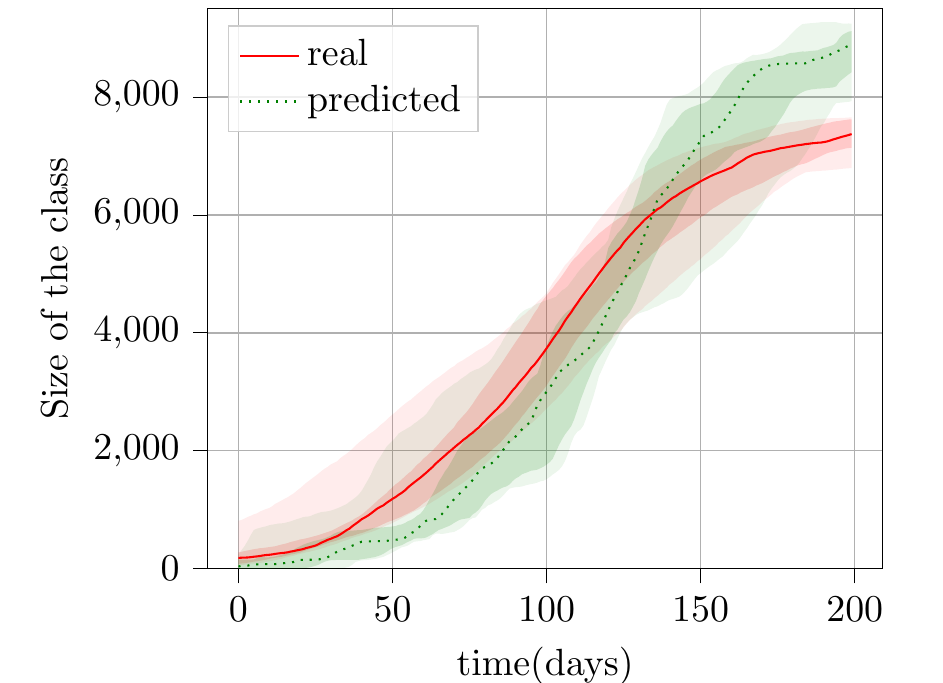}}
    \subfigure[S, 20\%]{\includegraphics[trim = {6cm 16.5cm 6cm 4.5cm}, clip, height = 0.225\columnwidth]{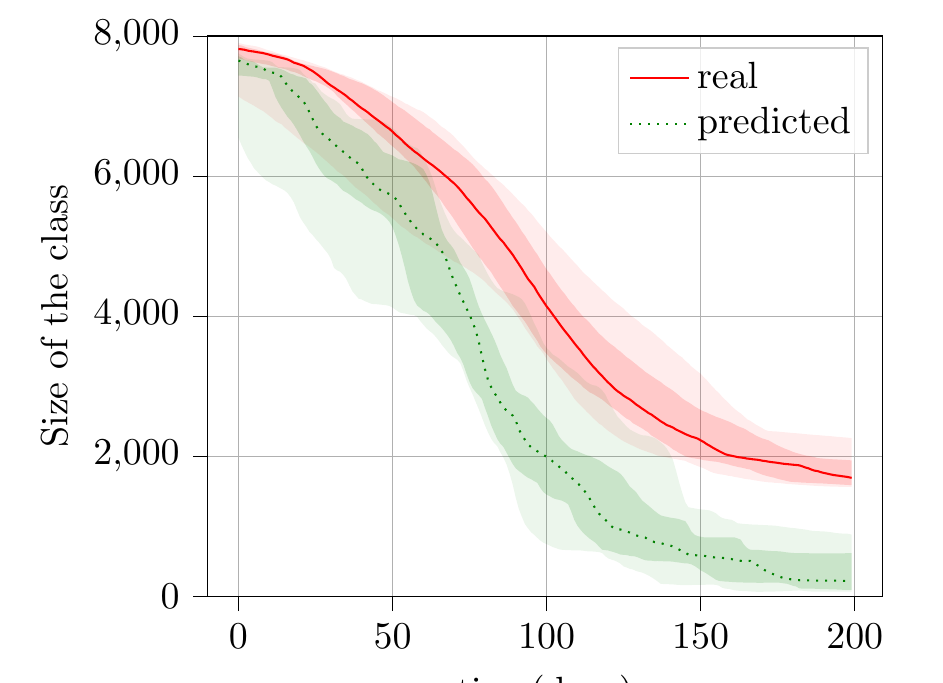}}
    \subfigure[E+I, 20\%]{\includegraphics[trim = {6cm 16.5cm 6cm 4.5cm}, clip, height = 0.225\columnwidth]{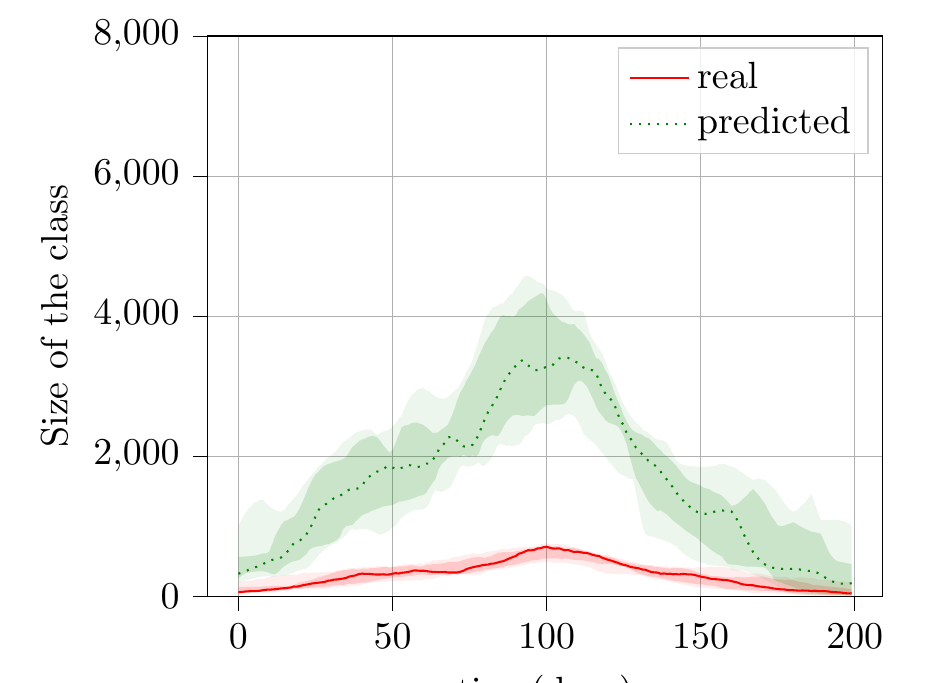}}
    \subfigure[R+D, 20\%]{\includegraphics[trim = {6cm 16.5cm 6cm 4.5cm}, clip, height = 0.225\columnwidth]{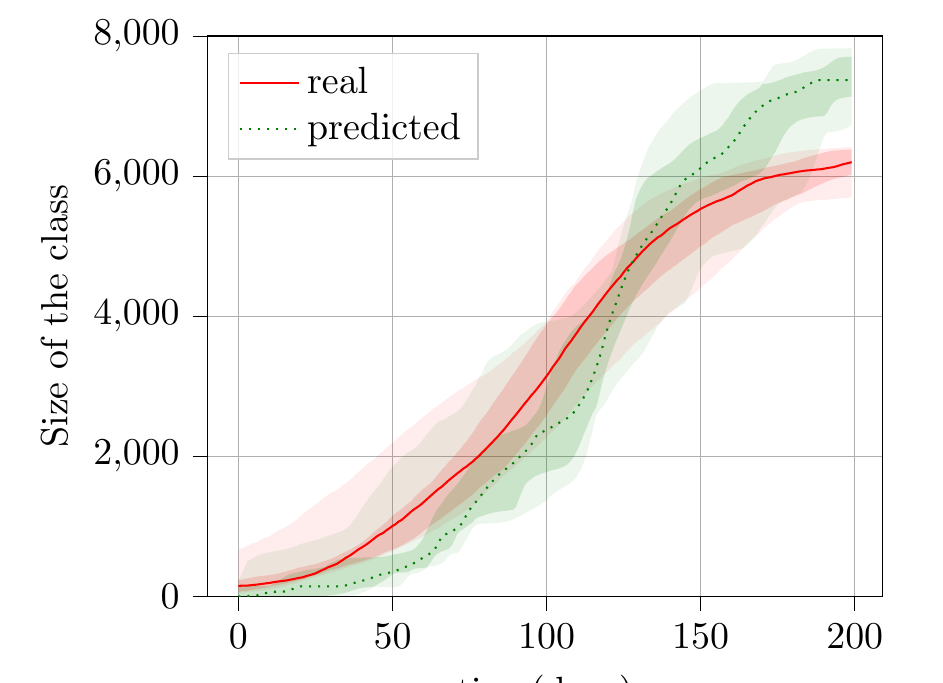}}
    \caption{Evolution of overall statistics associated to the epidemics, when evolving on a toy model inspired by the Boston and Cambridge (MA,USA) areas.
    We show the amount of number of susceptibles in Panels (a,d), the infected and exposed in Panels (b,e), the recovered or dead in Panels (c,f).
    Actual evolutions are in red, while estimations are in green. The solid lines represent the mean, while the translucent areas the variance.
    Training and testing sets are made of realizations produced by simulating the epidemic spreading on random social networks of 500 subjects.
    In Panels (a-c) only $5\%$ of subjects is tested at any time, while in Panels (d-f) this number reaches $20\%$.
    }
    \label{fig:results_boston}
\end{figure}

We test the effectiveness of the proposed approach by collecting a testing set made of $10$ realizations. Each realization is an evolution of the epidemic on a different and randomly generated social network (following the same statistical characteristics of the testing set). As for scenario 1, also here we test the case of size of $\mathcal{M}$ (i.e. tested subjects) being $5\%$, $10\%$, or $20\%$ of the total population. Results are shown in Tab. \ref{tab:scenario_boston} and Fig. \ref{fig:results_boston}. 
Although lower in the easier scenario 1, the accuracy is consistently good across classes and conditions. Yet, the accuracy of $S$ is a bit lower than before, and the precision of $E+I$ is very low. This is because the neural architecture tends to wrongly label a number of nodes which are susceptible as infected.
Yet, it is important to underlie here that the neural network is working with a quite small amount of information on the spreading of infected subjects. Indeed, at its pick $E + I$ is less than the $10\%$ of the population, which with $10\%$ of measures means that the algorithm can rely on the knowledge of $10^2$ infected nodes.
This behavior is also evident in Fig. \ref{fig:results_boston}, where the total number of susceptible subjects is higher than estimated, and vice versa the infected subjects are lower than the neural network thinks.
It is again important to stress that the proposed algorithm is optimized to estimate the distribution of subjects rather than the total size of each class, which should therefore be regarded as a secondary index.
It is also interesting that the algorithm rarely does the opposite error, i.e. classifying $S$ as $E + I$. The precision of $S$ is indeed above $97\%$. Although not explicitly forced in training phase, this behavior makes very much sense in the practice since it is better to isolate healthy subjects than not to act on infected ones.

\begin{table}[t]\label{tab:scenario_boston}
\caption{Accuracy (Ac.) and precision (Pr.) of the classification for Scenario 2 (Boston), evaluated only on the nodes which are not in $\mathcal{M}$. Three levels of supervision are considered - i.e., number of nodes of $V$ which are in $\mathcal{M}$ as well.}
\begin{center}
\begin{tabular}{ |c|c|c|c|c|c|c|c| } 
\hline
 & Ac., $5\%$ & Pr., $5\%$ & Ac., $10\%$ & Pr., $10\%$ & Ac., $20\%$ & Pr., $20\%$ \\
\hline
$S$        & 0.67 & 0.97 & 0.67 & 0.98 & 0.67 & 0.98 \\ 
$E \!+\! I$& 0.75 & 0.14 & 0.78 & 0.14 & 0.80 & 0.15 \\ 
$R \!+\! D$& 0.78 & 0.79 & 0.79 & 0.79 & 0.80 & 0.80 \\ 
All        & 0.72 & -    & 0.72 & -    & 0.73 & -   \\ 
\hline
\end{tabular}
\end{center}
\end{table}

\section{Discussion and Conclusions}

%
With this work we investigated the use of Graph Neural Networks to develop state observes for epidemics evolving on social networks.
The results are promising. The neural architecture can approximate the overall state with an accuracy which is always above the $70\%$, even when the sample space is as small as $5\%$ of the total population. 

Nonetheless, there are several directions towards which our results may be improved which we aim at investigating in the future. First, future work will be devoted to adding explicit dynamic reasoning withing the neural network, for example by introducing recurrent layers \citep{liang2016semantic}. This should help boost the capability of the neural network of discerning between exposed and infected, and between infected and recovered (or dead). Indeed, these transitions are essentially time dependent and can be extracted from associating an internal dynamics to the initial recognition that a node entered in the exposed state.
Yet, it is worth mentioning that stacking LSTMs layers in between the GNNs did not produce a statistically relevant increasing of the network performance and as such has not been included in the present work. Similarly the use of attention mechanisms \citep{velivckovic2017graph} have been tested but not included due to the negligible increment of performance that they resulted into.
Finally, we believe that a very important assumption to be relaxed is the full knowledge of the social network (see Sec. \ref{sec:goal}). Several algorithms are being proposed that can extract the social structure from GPS localization and other mobility information provided by contact tracing apps \citep{ferretti2020quantifying,cheng2020contact}.
Data driven methods can then possibly be used to infer the graph topology itself \citep{segarra2017network,giannakis2018topology}.

\section*{Conflicts of Interest}
The authors declare that they have no conflict of interest.

\section*{Funding}
This work is supported by the TU Delft CoVid-19 response fund, and by the Leverhulme Trust through the Research Project (Grant number RPG2017-370).

\bibliographystyle{spbasic_updated}      
\bibliography{sample_library}   

\end{document}